# Metamorphic Testing for Smart Contract Vulnerabilities Detection

**Jiahao Li**

School of Computer Science, Wuhan University, Wuhan 430072, China;
jiahao_li@whu.edu.cn

**Abstract:** Despite the rapid growth of smart contracts, they are suffering numerous security vulnerabilities due to the absence of reliable development and testing. In this article, we apply the metamorphic testing technique to detect smart contract vulnerabilities. Based on the anomalies we observed in vulnerable smart contracts, we define five metamorphic relations to detect abnormal gas consumption and account interaction inconsistency of the target smart contract. Through dynamically executing transactions and checking the final violation of metamorphic relations, we determine whether a smart contract is vulnerable. We evaluate our approach on a benchmark of 67 manually annotated smart contracts. The experimental results show that our approach achieves a higher detection rate (TPR, true positive rate) with a lower misreport rate (FDR, false discovery rate) than the other three state-of-the-art tools. These results further suggest that metamorphic testing is a promising method for detecting smart contract vulnerabilities.

## 1. Introduction

Smart contracts are self-executing programs that currently facilitate a variety of online decentralized finance transactions. A smart contract is a small computer program stored on the blockchain that converts traditional agreements into digital counterparts and automatically executes when specific conditions are satisfied [1]. Typically, smart contracts are a set of codes mainly written by Solidity [2] that execute on top of Ethereum [3], which is one of the most prominent blockchain platforms supporting smart contracts. All transactions pertaining to smart contracts are persistently and transparently stored on the blockchain without the need for trusted third-party verification [4]. Due to this benefit, smart contracts are utilized in numerous industries, e.g., decentralized finance, insurance, product tracking, and banking [5].

Although smart contracts are widely used for online commercial transactions, they have also been vulnerable to malicious attacks in the past due to defective or unreliable codes in smart contracts. Unfortunately, the absence of a reliable development and testing process has facilitated those malicious attacks resulting in significant financial losses. One infamous example is the "DAO" attack which led to the Ethereum main chain hard forked and stole over 3.5 million Ether resulting in about $60 million USD in losses at the time from the "DAO" contract [6].

These malicious attacks have prompted researchers to develop different methods for detecting smart contract vulnerabilities [6–15]. However, existing techniques are insufficient due to their low detection rate and high false alarm rate. The reason is that existing techniques mainly rely on predefined vulnerability patterns to detect vulnerable smart contracts [16]. Thus, the effectiveness of the tools relies heavily on the quality of the predefined vulnerability patterns. Besides, most patterns are defined by code static analysis without observing the actual effects of the transactions, resulting in missed and misreported vulnerabilities [17]. For example, in a contract misreported by Mythril [14] in Fig 8, the contract uses low-level *call()* to call another method *receiveApproval()*, and checks the status of the *call()* in line 4. It is the proper way to handle the *call()*, but Mythril considers it insecure as the contract violates the pattern "*call(s) without wrapping require()*". Another example in Fig 5, the contract indeed contains a reentrancy vulnerability that enables malicious contracts to steal Ether from it, but Slither [10] and Mythril miss it due to the violation of "*state variables written after the call(s)*" (more details in Section 6.2).

To address the above problems, we dynamically execute multiple transactions instead of performing static analysis and observe the actual gas consumption, transaction status and contract state or balance change in order to detect vulnerabilities. We observe that most of the vulnerable smart contracts are insufficient to counter malicious smart contracts' intentional manipulation of gas allocation and account switching. When facing such exploitation, the transactions of vulnerable contracts typically exhibit abnormal gas consumption and account interaction inconsistency (more details in Section 3). As a result, the contract state or balance becomes inconsistent with the expectation, indicating a potential vulnerability.

Based on the above insight, we propose to adopt metamorphic testing (MT) to detect abnormal gas consumption and account interaction inconsistency at the transaction level. To achieve this, we define several metamorphic relations (MRs) to encode the above two abnormal scenarios and use them as test oracles to detect vulnerable smart contracts. MT is a property-based testing technique [18], which is used for alleviating the oracle problem [19] of software testing. The central component of MT is MRs, which encode the necessary properties of the target program in relation to multiple inputs and their expected outputs. In recent years, MT has been extended to a wide range of software activities, such as software validation [20–22], fault localization [23], AI system testing [24] and QA system assessing [25,26].

In this study, we define a total of five MRs to identify abnormal scenarios at contract transaction runtime. Transactions that violate any of the MRs indicate that the contract is vulnerable. We evaluate our approach on a benchmark of 67 manually annotated smart contracts. Compared with three state-of-the-art tools, ContractFuzzer, Slither and Mythril, we find that ContractFuzzer misreports 29/67 (43.28%) false vulnerable smart contracts (i.e., false positive), Slither and Mythril both missed 8/38 (21.05%) true vulnerable smart contracts (i.e., false negative), while our approach achieved the strongest detection ability with the fewest misreports. In this paper, we make the following novel contributions:

- To the best of our knowledge, we propose the first work applying the technique of metamorphic testing to detect security vulnerabilities of smart contracts on the Ethereum platform.
- We define five MRs by considering gas allocation and account switching, which can be used to detect a wide range of vulnerabilities, such as gasless send, reentrancy and exception disorder.
- We evaluate our approach on 67 manually annotated smart contracts reported in other studies and demonstrate its feasibility and effectiveness in detecting vulnerable contracts. Meanwhile, we also analyze the reason why other tools create false alarms and miss vulnerabilities.

The remainder of the paper is organized as follows. Section 2 provides the necessary background about the smart contract and metamorphic testing. Section 3 introduces our observations on vulnerable smart contracts, and Section 4 clarifies the overall approach and presents a list of MRs identified for smart contracts. Section 5 illustrates our experimental setup. The evaluation results on real smart contracts are shown in Section 6. Finally, we discuss related work and conclude in Sections 7 and 8, respectively.

## 2. Preliminaries

### 2.1. Blockchain and Smart Contract

The blockchain is a distributed, immutable ledger designed to facilitate the recording of transactions and the tracking of assets within a decentralized network maintained by self-governing miners [4]. In the blockchain network, each block consists of a group of transactions that are verified and executed by miners through different consensus protocols (e.g., proof-of-work [27], proof-of-stake [28]). Once a verified block has been successfully appended to the blockchain, no previous blocks can be reverted or tampered unless an attacker controls more than half of all (at least 51%) miners, which seems impossible. The immutability and decentralized feature of blockchain makes it suitable for many applications, such as insurance, product tracking, banking and decentralized finance transactions.

Smart contracts are one of the most successful applications of blockchain technology. A smart contract is a self-executing computer program running on Ethereum [29], which convert traditional agreements into digital counterpart and automatically execute when specific conditions are met. Leveraging the Turing-complete Ethereum Virtual Machine (EVM), a smart contract can be created, deployed and run at a specific *address* on the Ethereum, providing public interfaces and fields for external access [30]. Moreover, a special field called *balance* stores the number of cryptocurrencies owned by this contract.

Transactions are cryptographically signed instructions from accounts, which change the state of the EVM [31]. On the Ethereum network, a transaction is a message call from a source to a target address. Transactions are mainly used to transfer cryptocurrencies from one account to another, deploy a new contract to a new address and invoke the functions of a deployed contract. As transactions change the state of the EVM, an execution fee called "gas" needs to be paid for the network (more details in Section 3.1). All external transactions are initiated by external users. In addition to external transactions, an on-chain contract may invoke another on-chain contract through internal transactions (more details in Section 3.2).

*2.2. Metamorphic testing*

Metamorphic testing (MT) is a property-based testing technique [18], which is used for alleviating the oracle problem [19]. MT encodes necessary properties, so called metamorphic relations (MRs), of the target program in relation to multiple inputs and their expected outputs. More specifically, an MR consists of two parts of constraint relations, one for constructing follow-up inputs from source inputs and the other for defining the expected relationship between follow-up outputs and source outputs. Consider testing the *sin(x)* program as an example for illustration, an MR for *sin(x)* can be "$sin(\pi - x) = sin(x)$, suppose the source input is an arbitrary angle $x$, and the follow-up input is $\pi - x$, as a result, the source output and follow-up output is expected to be equal".

Generally, there is no specific method to guide how to define metamorphic relations. Metamorphic relations can be defined in a variety of ways, such as analyzing the program's requirements, the source codes, the output behaviors or execution status, etc. Once an MR has been identified, MT will follow a standard testing procedure as below. The first step is to generate a series of source inputs, and then the next step is to construct follow-up inputs according to the MR's input constraint relation. After that, the program will be respectively executed with the source and follow-up inputs, and the source and follow-up outputs will be recorded. Finally, the relationship between source and follow-up outputs will be examined by the MR's output constraint relation to check whether or not the MR is violated. By checking the final MRs violation, we can determine whether or not the program is vulnerable.

The biggest difference between traditional testing techniques and metamorphic testing is that MT only examines the violation of MRs on groups of source and follow-up outputs rather than checking the correctness of outputs. Due to this benefit, MT has been extended to a wide range of software activities, such as software validation [20–22], fault localization [23], AI system testing [24] and QA system assessing [25,26].

**3. Observations on Vulnerable Smart Contracts**

In this section, we first analyze the expected gas consumption patterns and account interaction consistency when executing transactions of bug-free smart contracts, then describe our observations of some anomalies when executing transactions of vulnerable smart contracts.

*3.1 Abnormal gas consumption scenarios*

**Gas limit, gas price and transaction fee.** In Ethereum, every transaction has a specified amount of gas to be consumed for execution [32]. The *gas limit* determines the maximum amount of computational effort that can be used to execute a transaction, while the *gas price* is the amount of Ether that the transaction sender is willing to pay for each unit of gas consumed. The issuer of a transaction sets both *gas limit* and *gas price*. If the execution of a transaction requires consuming more gas than that specified by the *gas limit* parameter, such a transaction fails with an out-of-gas exception and gets rolled back [32]. The actual *transaction fee* depends on the final amount of *gas cost* and defines as *gas cost × gas price*. The *transaction fee* is also paid for failed transactions, including those with out-of-gas exceptions.

**Intrinsic gas cost.** From a low-level perspective, the intrinsic gas cost of a transaction depends on the number and type of bytecode operations executed during runtime. The gas cost of all bytecode operations is described in the Ethereum yellow paper [30]. In addition, a transaction may contain internal transactions. Thus, the total gas cost of a transaction is equal to the sum of the gas cost of all the instructions and internal transactions [33] during runtime. Formally, the gas cost function *GC* of a transaction can be defined as

$$GC = \sum_{inst \in insts} GC_{OPCODE_{inst}} + \sum_{t \in inter\_trans} GC_t, \quad (1)$$

where $insts = (inst_1, inst_2, inst_3, ...)$ is the set of instructions in the execution path. $GC_{OPCODE_{inst}}$ is the gas cost of the bytecode operations in $inst$. $inter\_trans = (inter\_tran_1, inter\_tran_2, inter\_tran_3, ...)$ is the set of potential internal transactions. $GC_t$ is the gas cost of the corresponding internal transaction $t$. For more technical details about the definition of the gas cost formula, we refer the reader to the Ethereum yellow paper [30].

**Gas consumption scenarios.** Given transaction $tran$ of a smart contract C, $GC_{tran}$ represents its intrinsic gas cost calculated by Equation 1, we define $Gas_{consumption}$ as the actual gas cost of the transaction. Set $Gas_{limit}$ as the upper bound of gas allocated to $tran$ and $\sigma_{tran} \in \{success, failure\}$ represents the execution status of $tran$. Theoretically speaking, if C is a bug-free smart contract, we can observe the following gas consumption scenarios when executing the same transaction with different gas limit settings:

- If $Gas_{limit} \geq GC_{tran}$, then $\sigma_{tran} = success$ and $Gas_{consumption} \equiv GC_{tran}$. That means if the gas limit is equal to or higher than the amount of gas required to execute the transaction, then the transaction will successfully execute [34]. Moreover, the actual gas consumption is constantly equal to the intrinsic gas cost, no matter how the gas limit changes.
- If $Gas_{limit} < GC_{tran}$, then $\sigma_{tran} = failure$ and $Gas_{consumption} \equiv Gas_{limit}$. In other words, if the gas limit is less than what is needed to execute the transaction, then the transaction fails with an out-of-gas exception [34]. As fees are also paid for failed transactions, thus the actual gas consumption constantly equals the gas allocated.

**Abnormal gas consumption scenarios.** However, not all transactions meet the above scenarios. In practice, if C contains some flaws, we can find some abnormal gas consumption scenarios:

- Assume $Gas_{limit} > GC_{tran}$, we expect $\sigma_{tran} = success$ and $Gas_{consumption} \equiv GC_{tra}$, but we find $Gas_{consumption} > GC_{tran}$. As gas cost depends on the low-level bytecode instructions, this gas consumption pattern implies that **extra but not expected** instructions are executed. As a result, the actual gas consumption is larger than the intrinsic gas cost.
- Assume $Gas_{threshold} \leq Gas_{limit} < GC_{tran}$, we expect $\sigma_{tran} = failure$, but we find $\sigma_{tran} = success$, where $Gas_{threshold}$ is the minimum threshold that enables the transaction to succeed without an out-of-gas exception. This pattern implies that ***necessary***

*and expected* instructions are not executed (we can further deduce that the unexecuted instructions gas cost is $GC_{tran} - Gas_{limit}$). As a result, a transaction is executed incorrectly.

*3.2 Account interaction inconsistency*

**Account types.** There are two types of accounts, externally-owned account (EOA) and contract account (CA), in Ethereum. The former is controlled by anyone with private keys, and the latter is controlled by code [35]. Both account types have the ability to: 1) receive, hold and send ETH; 2) interact with deployed smart contracts. The main difference between the two accounts is that the contract account uses a particular anonymous function, i.e., the fallback function, to receive Ether.

The fallback function is an anonymous external function in a smart contract with no input and output parameters. It will be executed if the invoked function does not exist in the contract [2]. Besides, the fallback function will be executed automatically when other accounts send Ether to the contract. In some cases, the fallback function can only rely on 2300 gas to execute when the sender contacts use *send()* or *transfer()* to send Ether [36].

In this work, we further categorize CA into four types based on the code operations performed within its fallback function. 1) **CAO** is a contract account with an empty fallback function. It means no code snippets are embedded in the fallback function. From a functional level perspective, a CAO is equivalent to an EOA; 2) **CAH** is a contract account with a heavy gas consumption fallback function that throws no exceptions. By inserting the code snippets containing heavy gas cost operations (such as an SSTORE operation which costs 5000 units gas), we can construct the fallback function that consumes more than 2300 gas; 3) **CAR**, a contract account with recursive call fallback function. The code snippets in its fallback function make a recursive call to the target contract, aiming to trigger reentrancy; 4) **CAE** is a contract account that contains explicit exceptions. We construct CAE by injecting a throw statement (e.g., *revert()*) in the fallback function to trigger exceptions intentionally.

**Account interaction consistency scenarios.** As both types of accounts, EOA and CA, can interact with a smart contract, the code logic in a bug-free smart contract should take into consideration the uncertain types of interacting accounts and interact consistently for any interacting account. Let us go further to explain and comprehend it. Suppose a smart contract C is an Ether holder to which different accounts can deposit and withdraw Ethers. External account A and contract account B are authorized users of C. Both A and B can interact with C via an external or internal transaction. In addition, A and B have equal eligibility, the difference is that B uses a fallback function to receive ethers. Given two withdraw transactions $A.tran$ and $B.tran$ with the same *amount* to C, $\sigma_{A\ or\ B.tran} \in \{success, failure\}$ represents the execution status of $A\ or\ B.tran$ and $\mu_{A\ or\ B}$ represents the balance change of $A\ or\ B$. Let $B.fallback \in M_{fallback}$ as the fallback function of B, where $M_{fallback}$ is a set of fallback functions with different injected code snippets. For a bug-free C, we can observe the following interaction consistency scenario when executing the same amount transaction with different interacting accounts:

- Given $B.fallback \in M_{fallback}$, and $B.fallback$ has no exceptions, if $\sigma_{A.tran} = success \wedge \sigma_{B.tran} = success$, then $\mu_A \equiv \mu_B$. In this case, the actual balance changes of A and B keep consistent when we switch the interacting account type from EOA to CAs with different $fallback$. That implies that the transfer code logic in C can correctly handle different types of interaction accounts and keep the results consistent. In further, C is robust to cope with the extra instructions in $B.fallback$.
- Given $B.fallback \in M_{fallback}$, if $B.fallback$ has exceptions, then $\sigma_{B.tran} = failure \wedge \mu_B \equiv 0$. In this scenario, $B.tran$'s execution status $\sigma_{B.tran}$ is always failure and the side effects of $B.tran$ need to be reverted, making B's balance change $\mu_B$ is always null.

**Account interaction inconsistency scenarios.** However, we found interaction inconsistency scenarios when C is flawed:

- Given $B.fallback \in M_{fallback}$ and $B.fallback$ has no exceptions, we find $\mu_A \neq \mu_B$ when $\sigma_{A.tran} = success \land \sigma_{B.tran} = success$. This pattern shows that the execution result of $A.tran$ and $B.tran$ is inconsistent despite the transactions eventually executing successfully, suggesting that $B.fallback$ may introduce some effects but C does not handle them correctly.
- Given $B.fallback \in M_{fallback}$ and $B.fallback$ has exceptions, we expect $\sigma_{B.tran} = failure \land \mu_B \equiv 0$, but we find $\sigma_{B.tran} = success \land \mu_B \equiv 0$. This pattern shows that the exceptions in $B.fallback$ are not well handled by C, resulting in an error transaction execution status.

## 4. Metamorphic Testing for Smart Contract Vulnerabilities Detection

This section first describes the overview framework of our approach, then presents the details of MRs derived from the observations in Section 3, and also provides illustrative examples to explain how previously reported vulnerabilities can be detected by our MRs.

*4.1 Overview Framework*

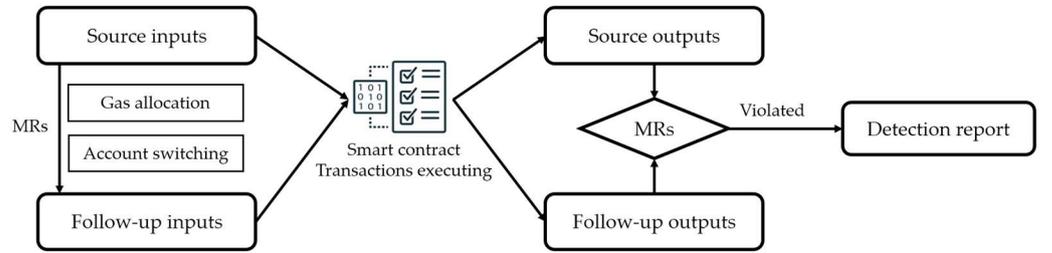

**Figure 1.** Overview of applying metamorphic testing to detect smart contract vulnerabilities

An overview of our approach is presented in Figure 1. Given a set of source inputs of a transaction and a group of metamorphic relations (MRs), our approach constructs the corresponding set of follow-up inputs according to the MRs' input constraint relation (e.g., gas allocation and account switching). Then, our approach respectively executes the transactions with source and follow-up inputs and records the corresponding source and follow-up outputs. Finally, the relationship between source and follow-up outputs will be examined by the MR's output constraint relation to check whether or not an MR is violated. By checking the final MRs violation, we can determine whether or not the smart contract is vulnerable.

*4.2 Metamorphic Relations*

In this work, we rely on dynamically executing transactions **in the same context** with different gas allocation and account switching. We can simply model a transaction as a tuple $tran := <A, G, E, \sigma, \delta, \mu>$, where $A$ is the account that interacts with the target smart contract, $G$ is the gas limit allocation to the transaction, and $E$ denotes the context of the transaction's execution; $\sigma, \delta, \mu$, respectively, represent the transaction's execution status, gas consumption and balance changes of $A$. Let $t_s$ and $t_f$ be a group of source and follow-up inputs of a transaction with respect to an MR, and let $O_s$ and $O_f$ be the corresponding source and follow-up outputs. Note that the outputs $O_s$ and $O_f$ may consist of one or more results from $\sigma, \delta, \mu$; we will provide a detailed explanation later.

The transaction's input consists of an interacting account, a gas limit allocation and an unchanged context. As such, we use $A_s$ and $G_s$ to denote the account and the gas allocation in $t_s$, and use $A_f$ and $G_f$ to denote the corresponding data in $t_f$. That is $t_s = tran<A_s, G_s, E>$ and $t_f = tran<A_f, G_f, E>$. Different MRs may operate on different input parameters of $t_s$ to construct $t_f$, leading to discrepancies between $t_s$ and $t_f$. According to this, we summarize MRs in Table 1 and explain them below.

- MR1.x has $t_s = tran < A, G_s, E >$ and $t_f = tran < A, G_f, E >$. That is, $t_s$ and $t_f$ in MR1.x are executed by the same account A within the same context E, but the different gas limit allocation $G_s$ and $G_f$. MR1.x operate on $G_s$ to construct $G_f$ and focus on detecting abnormal gas consumption patterns for the target smart contract.
- MR2.x has $t_s = tran < A_s, G, E >$ and $t_f = tran < A_f, G, E >$. That is, $t_s$ and $t_f$ in MR2.x are executed within the same gas limit allocation $G$ and the same context $E$, but different interacting account $A_s$ and $A_f$. MR2.x operate on $A_s$ to construct $A_f$ and focus on detecting account interaction inconsistency patterns for the target smart contract.

**Table 1.** Summary of metamorphic relations (MRs).

|  | Source and Follow-Up Inputs | Number of MRs |
| --- | --- | --- |
| MR 1.x | $t_s = tran < A, G_s, E >$, $t_f = tran < A, G_f, E >$ | 2 (MR1.1–MR1.2) |
| MR 2.x | $t_s = tran < A_s, G, E >$, $t_f = tran < A_f, G, E >$ | 3 (MR2.1–MR2.3) |

4.2.1 Gas allocation MR1.x.

MR1.x is designed to detect abnormal gas consumption patterns for the target smart contract. Different MRs alter $G_s$ to construct $G_f$ by mutating the gas limit allocation with respect to intrinsic gas cost (i.e., $GC_{tran}$ in Section 3.1) and also encode the relationship that is expected to be satisfied by $O_s$ and $O_f$.

- MR1.1 (Increasing gas allocation): Given $G_s \geq GC_{tran}$, $G_f$ is constructed by increasing the gas allocation value of $G_s$. After executing the source and follow-up transaction, the expected relationship between $O_s$ and $O_f$ is defined as $O_s(\sigma_s, \delta_s) = O_f(\sigma_f, \delta_f)$ $iff$ $\sigma_s = \sigma_f \wedge \delta_s = \delta_f$, where $\sigma_{s\,or\,f}$ and $\delta_{s\,or\,f}$ denote the transaction's execution status and gas consumption of $t_{s\,or\,f}$ respectively.
- MR1.2 (Reducing gas allocation): Given $G_s = GC_{tran}$, $G_f$ is constructed by reducing gas allocation value of $G_s$. After executing the source and follow-up transaction, the expected relationship is defined as $O_s(\sigma_s) \neq O_f(\sigma_f)$ $iff$ $\sigma_s \neq \sigma_f$.

4.2.2 Account switching MR2.x

MR2.x intends to detect account interaction inconsistency patterns for the target smart contract. MR2.x leverages mutating the fallback function to generate $A_f$ from $A_s$.

- MR2.1 (Switching EOA to CAH): Given an EOA $A_s$, $A_{temp}$ is an equivalent CAO to $A_s$. $A_f$ (i.e., a CAH contract account) is constructed by inserting heavy gas consumption operations to $A_{temp}$'s fallback function. After executing the source and follow-up transaction, the expected relationship between $O_s$ and $O_f$ is defined as $O_s(\sigma_s, \mu_s) = O_f(\sigma_f, \mu_f)$ $iff$ $\sigma_s = \sigma_f \wedge \mu_s = \mu_f$, where $\mu_{s\,or\,f}$ denotes the balance changes of $A_{s\,or\,f}$.
- MR2.2 (Switching EOA to CAR): Set $A_{temp}$ as an equivalent CAO to a given EOA $A_s$. $A_f$ (i.e., a CAR contract account) is constructed by embedding a recursive call to $A_{temp}$'s fallback function. After executing the source and follow-up transaction, the expected relationship between $O_s$ and $O_f$ is defined as $O_s(\sigma_s, \mu_s) = O_f(\sigma_f, \mu_f)$ $iff$ $\sigma_s = \sigma_f \wedge \mu_s = \mu_f$.
- MR2.3 (Switching EOA to CAE): Set $A_{temp}$ as an equivalent CAO to a given EOA $A_s$. $A_f$ (i.e., a CAE contract account) is constructed by embedding a throw statement to $A_{temp}$'s fallback function. After executing the source and follow-up transaction, the expected relationship between $O_s$ and $O_f$ is defined as $O_s(\sigma_s) \neq O_f(\sigma_f)$ $iff$ $\sigma_s \neq \sigma_f$.

*4.3 Detecting Vulnerabilities with MRs*

**Reentrancy.** The reentrancy vulnerability is caused by the fact that vulnerable contracts fail to properly prevent the potential recursive calls in malicious contracts' fallback function [37–39]. Consider the simplified "DAO" attack as an example. *SimpleDAO* in Fig

2 allows different types of accounts to withdraw Ether using *withdraw(amount)* in lines 6-10. If the withdrawer is an EOA or CAO, *withdraw(amount)* will function properly, and the withdrawing account's *balance* will be credited with *amount* Ether. However, the withdrawer can be a malicious contract account, such as *Attacker* in Fig 3. When *Attacker* uses line 10 in Fig 3 to withdraw Ether, it will execute lines 7-8 in Fig 2. Then, the fallback function (lines 12-15 in Fig 3) of *Attacker* will be invoked automatically, and line 14 in Fig 3 will execute lines 7-8 of Fig 2 again and thus make recursive calls. Since *SimpleDAO* lacks proper conditions to prevent recursive calls from *Attacker*, *Attacker* will receive more than *amount* Ether from *SimpleDAO*, thus violating MR2.2. Besides, as *Attacker* performs multiple recursive calls to *SimpleDAO*, the actual gas consumption of the transaction will vary with the actual number of recursive calls, which will violate MR1.1.

```solidity
1 contract SimpleDAO {
2     mapping(address => uint256) public balances;
3     function deposit(address to) public payable {
4         balances[to] = msg.value;
5     }
6     function withdraw(uint amount) public {
7         require(balances[msg.sender] >= amount);
8         msg.sender.call.value(amount)(); // exception disorder
9         balances[msg.sender] -= amount;
10    }
11    function withdraw_a(uint amount) public {
12        require(balances[msg.sender] >= amount);
13        balances[msg.sender] -= amount;
14        msg.sender.call.value(amount)(); // exception disorder
15    }
16    function withdraw_b(uint amount) public {
17        require(balances[msg.sender] >= amount);
18        msg.sender.send(amount); // gasless send
19        balances[msg.sender] -= amount;
20    }
21 }
```

**Figure 2.** A simplified DAO contract (victim).

**Exception Disorder.** The Exception disorder is caused by Solidity's inconsistent exception handling [37]. Solidity provides two ways of exception handling. Given a chain of all direct calls, the side effect of transactions will be reverted when an exception occurs. Given a chain of calls with low-level call methods, such as *call()*, *delegatecall()* and *send()*, the side effect of transactions will be reverted along the chain until it reaches the nearest low-level call method, at which point the low-level call method will return false. For example, line 14 in Fig 2 sends Ether to the account *msg.sender* by *call()*. Assume the account's fallback function costs 5000 units gas, if the gas provided by *call()* is more than 5000 units, both the transaction and the Ether transfer will succeed. If the gas provided by *call()* is less than 5000 units, the external call in line 14 will suffer an out-of-gas exception, the Ether transfer will fail but the transaction will succeed because the low-level call only returns false and does not revert. Thus, this transaction will violate MR2.1. Besides, if we keep reducing the gas allocation to *call()*, the transaction will also violate MR1.2.

```
 1 contract Attacker {
 2     SimpleDAO public dao;
 3     constructor(address addr) public {
 4         dao = SimpleDAO(addr);
 5     }
 6     function deposit(address to) public payable {
 7         dao.deposit.value(msg.value)(to);
 8     }
 9     function withdraw(uint amount) public {
10         dao.withdraw(amount);
11     }
12     function() payable public {
13         if (conditions pass)
14             dao.withdraw(...);//reentrancy
15     }
16 }
```

**Figure 3.** A reentrancy attack contract (attacker).

**Gasless Send.** The gasless send is because the gas allocation to *send()* is strictly limited to 2300 by the EVM [37]. If the receiver contract's fallback function has heavy gas cost larger than 2300, an out-of-gas exception will occur, resulting in a gasless send. If the such exception is not checked and propagated appropriately, the receiver will suffer a loss and the vulnerable contract can keep Ether wrongfully while seemingly innocent. For example, line 18 in Fig 2 sends Ether to account *msg.sender* by *send()*. Assume the account's fallback function costs $x$ units of gas, if $x$ is less than 2300, both the Ether transfer and the transaction will succeed. If $x$ is larger than 2300, the Ether transfer will fail but the transaction will succeed. Thus, this transaction will violate MR2.1.

## 5. Experiments Setup

*5.1 Datasets*

In the experiments, we chose ContractFuzzer [12], Slither [10], and Mythril [14] for comparison. ContractFuzzer is a dynamic analysis tool that first introduces fuzzing testing to detect vulnerabilities. Both Slither and Mythril are well-known static analysis tools. According to [40,41], Slither has a higher vulnerability detection rate, while Mythril has a higher precision. To evaluate our approach, we use a benchmark consisting of 67 manually annotated smart contracts as our experimental subjects. We first collect all the 67 contracts reported as vulnerable by ContractFuzzer into our dataset. Then we manually check all these contracts to determine whether they are vulnerable. Finally, 38 of them are verified to be vulnerable. Table 2 lists the details of the benchmark.

**Table 2.** Summary of the benchmark.

| Vulnerability | All | Verified |
| --- | --- | --- |
| Reentrancy | 13 | 11 |
| Gasless Send | 17 | 6 |
| Exception Disorder | 37 | 21 |
| Total | 67 | 38 |

*5.2 Smart Contract Deployment*

As mentioned in Section 4.2, we rely on dynamically executing multiple transactions **in the same context** to realize different MRs. Therefore, how to eliminate potential changes in the context of multiple transactions is essential to our work. To maintain the same execution context for multiple transactions, we reset the state of interacting contracts (the initial contract and the target contract) each time we execute a transaction to eliminate the effect of changes to EVM values. As different EVM values' changes may cost different gas unit, such as an SSTORE operation costs 20000 when the storage value is changed from zero to non-zero, but only 5000 when the storage value's zeroness remains unchanged or is set to zero [30]. Besides, we adopted an additional strategy that simultaneously eliminates the impact of potential context changes and expedites the testing experiment. This approach deploys multiple replications of smart contracts before executing multiple transactions. When a new transaction needs to be executed, we simply pick the unused contracts (the only difference between these duplicate contracts is the deployment address of the contract). All the contracts are deployed in Goerli Testnet [42] by Remix [43]. Goerli Testnet is a cross-client proof-of-stake (PoS) network and one of Ethereum's most popular testnets. Remix is a comprehensive smart contract development tool.

*5.3 Gas Estimating and Allocating*

**Intrinsic gas cost estimating.** In section 3, we mentioned that each EVM bytecode operation consumes a specific amount of gas. Theoretically, we can accurately calculate the intrinsic gas cost of a transaction by Equation 1, but in practice, due to the unpredictability of the code's execution path, it is impossible to predict gas cost beforehand. Fortunately, Ethereum provides a standard function *estimateGas()* to estimate the gas consumption of a transaction [44]. However, this function's estimate is inaccurate and it tends to underestimate the amount when a transaction contains internal transactions. Therefore, we use the following heuristic steps to estimate the intrinsic gas cost of a particular transaction.

1) We first use the standard tool *estimateGas()* to roughly estimate the intrinsic gas cost $GC_{tran}$ of a transaction, here we donate the estimated value as $GC_{estimate}$;
2) Then we set the gas limit $Gas_{limit}$ to $GC_{estimate}$, execute the transaction by an EOA account to check whether the transaction is executed successfully;
3) If the transaction executes successfully, we set $GC_{tran}$ to the actual gas consumption $Gas_{consumption}$;
4) If the transaction fails and throws an out-of-gas exception, we keep increasing the gas limit $Gas_{limit}$ and repeat step 2) until the transaction executes successfully, then set $GC_{tran}$ to the actual gas consumption $Gas_{consumption}$.

**Gas allocating.** To realize MR1.x, we need to manipulate the gas allowance to generate follow-up test cases. We use the following specific steps in the experiments.

1) Increasing gas allocation. In Ethereum, there is a block gas limit [45] for each block to limit the maximum number of transactions that can be packed in each block, which we denote as $Gas_{block}$. Thus, we set the gas increasing space to $(GC_{tran}, Gas_{block}]$, and set the increasing interval to at least $GC_{tran}$.
2) Reducing gas allocation. We set the gas reduction space to $[0, GC_{tran})$. Afterward, we divide it into *n* equal parts and reduce the gas allocation value by $GC_{tran}/n$ each time, here we set *n=1000*.

*5.4 Agent Contracts*

MR2.x intends to detect account interaction inconsistency for the target smart contract. Therefore, realizing MR2.x requires interactive calls between two CA contracts. As contract transactions are always initiated by an EOA account (i.e., the transaction sender is always an EOA account), to realize MR2.x, different agent contracts must be constructed to interact with the target contracts. Different agent contracts are described in Fig 4.

As shown in Fig 4, different agent contracts have the same *AgentCall* function but different fallback functions. An *AgentCall* function (lines 6-10) is designed to make a call to the target contract. The argument *contract_addr* records the target contract's address. The argument *msg_data* consists of the target function and the argument values passed to the function. The argument *contract_addr* and *msg_data* will be assigned to global variables *target_contract* and *call_msg_data* respectively in lines 7-8 so it can be used in other functions. Four different types of fallback functions are defined in lines 12-22 (Please note that we have combined four separate agent contracts into one contract to explain it in order to save space). A CAO agent has an empty fallback function in line 13. A CAH agent has a heavy gas cost fallback function, and line 15 implements this using a storage variable self-increment operation. A CAE agent with an exception throw statement in its fallback function is defined in line 17, while a CAR agent with a reentrancy call is defined in lines 19-22. A condition checker is used in line 20 to prevent endless reentrancy and out-of-gas exception. The recursive call in line 21 uses two global variables, *target_contract* and *call_msg_data*, to generate a reentrant attack scenario to try triggering reentrancy vulnerability.

```
1  contract Agent {
2    address public target_contract;
3    bytes public call_msg_data;
4    uint public storage_value;
5    ...
6    function AgentCall(address contract_addr, bytes msg_data) {
7      target_contract = contract_addr;
8      call_msg_data = msg_data;
9      target_contract.call(call_msg_data);
10   }
11   ...
12   // CAO agent
13   function() payable {}
14   // CAH agent
15   function() payable { storage_value++; }
16   // CAE agent
17   function() payable { revert(); }
18   // CAR agent
19   function() payable {
20     if (specified_condition pass)
21       target_contract.call(call_msg_data);
22   }
23 }
```

**Figure 4.** An illustration agent contract with different fallback functions

**6. Results and Analysis**

*6.1 Metrics*

To evaluate our approach, we collected several experimental measurements, such as true positives (TP), false negatives (FN) and false positives (FP). TP represents the number of contracts containing vulnerabilities correctly identified as vulnerable by the tool. FN represents the number of vulnerable contracts missed by the tool. FP indicates the number of contracts misreported as vulnerable by the tool. We define a vulnerability as correctly detected by our approach when any of the test cases in the groups of source and follow-up test cases violate the defined MRs. Two essential metrics, TPR and FDR, are computed using these measurements to evaluate the performance of the tools. TPR (true positive rate, also called recall) indicates the effectiveness of a tool in detecting actual vulnerabilities. A high TPR indicates that a tool effectively detects vulnerabilities and has a low rate of false negatives. FDR (false discovery rate) implies the misreported rate of a tool. A high FDR indicates that a tool is inaccurate in identifying actual vulnerabilities, as it is more likely to report false positives. Equations 2 and 3 give the definition of TPR and FDR. A tool that achieves higher TPR and lower FDR is considered to be better in terms of accurately identifying actual vulnerabilities and avoiding false positives.

$$TPR = \frac{TP}{TP + FN} \qquad (2)$$

$$FDR = \frac{FP}{TP + FP} \qquad (3)$$

*6.2 Effectiveness of MR*

Table 3 and Table 4 present the results of our experiments. Table 3 compares the overall performance of our method to that of three other tools. As the data shows, our approach and ContractFuzzer achieve higher TPR (100%) than Slither and Mythril. However, ContractFuzzer misreports 29 FPs, resulting in the highest FDR (43.28%) among the four tools. Slither and Mythril both detect 30 TPs with a TPR of 78.95%. Slither and our approach report 0 FPs, achieving the lowest FDR (0%), while Mythril misreports 3 FPs. Compared with the three state-of-the-art tools, our approach achieves the highest TPR and lowest FDR. More specifically, our approach can detect most vulnerabilities without any misreporting.

**Table 3.** The overall performance of the four tools.

| Tool | TP | FP | FN | TPR | FDR |
|---|---|---|---|---|---|
| ContractFuzzer | 38 | 29 | 0 | 100.00% | 43.28% |
| Slither | 30 | 0 | 8 | 78.95% | 0.00% |
| Mythril | 30 | 3 | 8 | 78.95% | 9.09% |
| MR | 38 | 0 | 0 | 100.00% | 0.00% |

**Table 4.** The performance of each tool on different vulnerability categories.

| Vulnerability | ContractFuzzer | Slither | Mythril | MR |
|---|---|---|---|---|
| Reentrancy | 13 (02) | 7 (0) | 5 (0) | 11 (0) |
| Gasless Send | 17 (11) | 6 (0) | 6 (0) | 6 (0) |
| Exception Disorder | 37 (16) | 17 (0) | 22 (3) | 21 (0) |
| Total | 67 (29) | 30 (0) | 33 (3) | 38 (0) |

Table 4 indicates the performance of each tool on different vulnerability categories. Each row in Table 4 represents a vulnerability category, and each cell presents the number of contracts reported as vulnerable by the tool. The number within parentheses presents the number of contracts misreported by the tool.

As is shown in Table 4, Slither, Mythril and our approach detect 7, 5 and 11 reentrancy vulnerabilities respectively, and none of them are misreported. We manually check the contracts that are omitted by Slither and Mythril. We find that both Slither and Mythril only rely on a code pattern "*state variables written after the call(s)*" to detect reentrancy. Contracts that do not satisfy the above pattern are not detected. For example, in Fig 5, line 5 indeed contains a reentrancy vulnerability that enables malicious contract to steal Ether from it, but this contract is omitted by Slither and Mythril due to the violation of above pattern. ContractFuzzer detects 13 reentrancy vulnerabilities. Out of these, 2 contracts are misreported. We manually checked the contract code and confirmed that they are falsely detected. We find that ContractFuzzer misreports reentrancy vulnerabilities because of omitting the preconditions before a potential reentrancy call. For example, a contract misreported by ContractFuzzer in Fig 6, in line 3, the contract defines a *msg.value* checker to prevent a reentrancy attack before transferring Ethers. When a reentrancy call occurs, the *msg.value* will be set to zero, thus the checker will fail and the transaction will be reverted.

```
1    event ET(address indexed _pd, uint _tkA, uint _etA);
2    function eT(address _pd, uint _tkA, uint _etA) returns (bool success) {
3        balances[msg.sender] = safeSub(balances[msg.sender], _tkA);
4        balances[_pd] = safeAdd(balances[_pd], _tkA);
5        if (!_pd.call.value(_etA)()) revert();
6        ET(_pd, _tkA, _etA);
7        return true;
8    }
```

**Figure 5.** A reentrancy vulnerability missed by Slither and Mythril.

```
1    function pay(address _addr, uint256 count) public payable {
2        assert(changeable==true);
3        assert(msg.value >= price*count);
4        if(!founder.call.value(price*count)() || !msg.sender.call.value(msg.value-price*count)()){
5            revert();
6        }
7        s.update(_addr,count);
8        Buy(msg.sender,count);
9    }
```

**Figure 6.** A misreported reentrancy vulnerability by ContractFuzzer.

For gasless send, Slither, Mythril and our approach detect all 6 gasless send vulnerabilities without misreporting any vulnerability. ContractFuzzer detects the most 17 gasless send vulnerabilities. However, 11 of them are misreported up to FDR (64.71%). We manually examined the code and confirmed that ContractFuzzer incorrectly identifies a contract using the *transfer()* function as vulnerable. The reason is that the *transfer()* function automatically reverts the contract state if there is insufficient gas when sending Ether. Fig 7 shows this false scenario.

```
1    function withdraw() public {
2        var balance = dividend(msg.sender);
3        payouts[msg.sender] += (int256) (balance * scaleFactor);
4        totalPayouts += (int256) (balance * scaleFactor);
5        contractBalance = sub(contractBalance, balance);
6        msg.sender.transfer(balance);
7    }
```

**Figure 7.** A misreported gasless send by ContractFuzzer.

```
1 function approveAndCall(address _spender, uint256 _value, bytes _extraData) returns (bool success) {
2     allowed[msg.sender][_spender] = _value;
3     Approval(msg.sender, _spender, _value);
4     if(!_spender.call(bytes4(bytes32(sha3("receiveApproval(address,uint256,address,bytes)"))),
5         msg.sender, _value, this, _extraData)) { return false; }
6     return true;
7 }
```

**Figure 8.** A misreported exception disorder by Mythril.

For exception disorder, Slither and our approach detect 17 and 21 exception disorder vulnerabilities respectively, and none of them are misreported. We further examine the code and find that the contracts omitted by Slither are defined as a type of "*functions that send Ether to arbitrary destinations*" vulnerability. Mythril reports 22 vulnerabilities with 3 FPs. We find that Mythril defines an external call without wrapping *require()* as vulnerable. For example, in a contract misreported by Mythril in Fig 8, the contract uses low-level *call()* to call another method *receiveApproval()* and checks the status of the *call()* in line 4. It is the proper way to handle *call()*, but Mythril considers it insecure. ContractFuzzer detects 37 exception disorder vulnerabilities. However, 16 of them are misreported up to FDR (43.24%). ContractFuzzer has so many FPs because its detection rules are comparable to those of Mythril. Besides, ContractFuzzer also incorrectly identifies contracts utilizing the *transfer()* function as vulnerable.

**Summary.** Our approach is effective in detecting vulnerabilities in smart contracts. The comparative experiment with three state-of-the-art tools shows that our approach gets the best performance with higher TPR and lower FDR. More specifically, compared with ContractFuzzer which achieves high TPR, our approach reports no false alarms. Compared with Slither and Mythril which have low FDR, our approach can detect 26.67% more vulnerabilities.

**7. Related Work**

A number of smart contract vulnerability detection methods have been proposed, and they fall into two categories: static analysis and dynamic analysis.

*7.1 Static Analysis*

**Code analysis.** SmartCheck [11] is an extensible static analysis tool that translates source code into an XML-based intermediate representation (IR) and uses XPath queries on IR to check whether a smart contract violates the predefined patterns. Securify [46] utilizes domain-specific information to define compliance and violation patterns and uses those patterns to detect vulnerable smart contracts. Slither [10] is an open-source static analysis framework for smart contracts, which leverages an intermediate representation SlithIR to detect vulnerabilities.

**Symbolic execution.** Oyente [13] is a pioneer work that first applies symbolic execution to detect smart contract vulnerabilities. It defines four types of vulnerable patterns and uses symbolic execution to examine the violation of these patterns. Maian [47] and Osiris [48] are extensions of Oyente in that they can detect more vulnerability categories. Teether [15] identifies four critical EVM instructions paths to guide safety transfer and searches for these critical paths in a contract's control flow graph to detect a vulnerable smart contract. Mythril [14] combines symbolic execution with SMT solving and taint analysis to detect vulnerable smart contracts. DefectChecker [49] analyzes smart contracts' bytecode using symbolic execution and utilizes eight predefined rules to detect vulnerabilities. Vulpedia [17] proposes to use smart contracts' abstract vulnerable signatures to detect four types of vulnerabilities.

Most of the above static analysis methods have advantages regarding analysis time but still suffer from miss and misreport scenarios due to the quality of predefined detection rules. Our approach differs from these methods in that our approach dynamically executes real transactions of smart contracts; thus, the detected contracts are guaranteed to be vulnerable.

*7.2 Dynamic analysis*

**Fuzzing testing.** ContractFuzzer [12] is a pioneer work that first applies fuzzing testing to detect vulnerable smart contracts. It uses seven predefined vulnerable patterns to guide the fuzzing testing procedure. GasFuzzer [50] is an extended version of ContractFuzzer, which specifically adds gas mutation to the fuzzing testing procedure. ReGuard [39] and ReDefinder [38] are two fuzzing-based methods that specifically detect reentrancy vulnerabilities. They both encode reentrancy vulnerabilities into several call patterns. Sfuzz [9] proposes a feedback-based fuzzer. It uses the detection oracle from ContractFuzzer. ContraMaster [16] combines fuzzing testing with mutating the transaction sequences to detect vulnerable smart contracts. It defines two test oracles to capture transaction and balance invariant.

Our approach differs from these dynamic analysis methods in that our approach uses transaction-level test oracles, while other methods use syntax-level test oracles. As the syntax level test oracles cover a limited and fixed set of vulnerabilities, these dynamic methods usually suffer from miss and misreport scenarios. Besides, our approach only checks the relationship among different outputs rather than checking the correctness of different outputs.

## 8. Conclusions

In this paper, we apply metamorphic testing (MT) to detect vulnerabilities in smart contracts. Instead of the specific syntax vulnerability detection patterns or oracles used in previous work, we identify five general-purpose metamorphic relations (MRs) to detect vulnerabilities. The experiments on 67 manually checked contracts show that our proposed MRs can achieve the highest TPR and lowest FDR. More specifically, our approach can detect most vulnerabilities without any misreporting compared with three state-of-the-art tools. These results further suggest that metamorphic testing is a promising method for detecting smart contract vulnerabilities.